\documentclass[%
 preprint,
 superscriptaddress,
 amsmath,amssymb,
 aps,
]{revtex4-2}

\usepackage{graphicx}
\usepackage{epstopdf}
\usepackage{gensymb}
\usepackage{graphicx}
\usepackage{dcolumn}
\usepackage{bm}

\begin{document}


\title{Molecular mechanisms of self-mated hydrogel friction}

\author{Jan Mees}
\affiliation{Department of Microsystems Engineering, University of Freiburg, Georges-K\"ohler-Allee 103, 79110 Freiburg, Germany}
\affiliation{Cluster of Excellence livMatS, Freiburg Center for Interactive Materials and Bioinspired Technologies, University of Freiburg, Georges-K\"ohler-Allee 105, 79110 Freiburg, Germany}
\author{Rok Simi\v{c}} 
\affiliation{Laboratory for Surface Science and Technology, Department of Materials, ETH Zürich, Vladimir-Prelog-Weg 1-5/10, 8093 Zürich, Switzerland}
\author{Thomas C. O'Connor} 
\affiliation{Department of Materials Science and Engineering, Carnegie Mellon University, 4309 Wean Hall, Pittsburgh, Pennsylvania 15213, USA}
\author{Nicholas D. Spencer} 
\affiliation{Laboratory for Surface Science and Technology, Department of Materials, ETH Zürich, Vladimir-Prelog-Weg 1-5/10, 8093 Zürich, Switzerland}
\author{Lars Pastewka} 
\email{lars.pastewka@imtek.uni-freiburg.de}
\affiliation{Department of Microsystems Engineering, University of Freiburg, Georges-K\"ohler-Allee 103, 79110 Freiburg, Germany}
\affiliation{Cluster of Excellence livMatS, Freiburg Center for Interactive Materials and Bioinspired Technologies, University of Freiburg, Georges-K\"ohler-Allee 105, 79110 Freiburg, Germany}

\begin{abstract}
Self-mated hydrogel contacts show extremely small friction coefficients at low loads but a distinct velocity dependence.
Here we combine mesoscopic simulations and experiments to test the polymer-relaxation hypothesis for this velocity dependence, where a velocity-dependent regime emerges when the perturbation of interfacial polymer chains occurs faster than their relaxation at high velocity.
Our simulations reproduce the experimental findings, with speed-independent friction at low velocity, followed by a friction coefficient that rises with velocity to some power of order unity.
We show that the velocity-dependent regime is characterized by reorientation and stretching of polymer chains in the direction of shear, leading to an entropic stress that can be quantitatively related to the shear response.
The detailed exponent of the power law in the velocity dependent regime depends on how chains interact: We observe a power close to $1/2$ for chains that can stretch, while pure reorientation leads to a power of unity.
Our simulations quantitatively match experiments and show that the velocity dependence of hydrogel friction at low loads can be firmly traced back to the morphology of near-surface chains.
\end{abstract}


\maketitle


\section*{Introduction}

Hydrogels are chemically crosslinked, hydrophilic polymer networks that swell in water or aqueous solvents. They are responsible for the low friction coefficients found in some biological systems, such as cartilage or the cornea~\cite{gong2006}. Due to their tribological properties, hydrogels have been applied in soft contact lenses~\cite{rennie2005,roba2011,dunn2013} and as synthetic articular cartilage materials~\cite{bray1973,corkhill1990,milner2018}. 

Self-mated hydrogel-on-hydrogel frictional contacts, also known as Gemini hydrogels, have low friction coefficients at low sliding velocities~\cite{dunn2014}. It has been shown that the polymerization conditions define a hydrogel's degree of crosslinking close to the surface~\cite{meier2019,simic2021}. Hydrogels whose polymerization close to the surface is inhibited by the presence of oxygen, exhibit pronounced crosslinker gradients and sparse surfaces. Conversely, the lack of oxygen during synthesis leads to a homogeneously crosslinked hydrogel~\cite{meier2019,simic2021}. Nonetheless, no polymer network is defect free and this will introduce dangling chains and loops on the surface~\cite{zhong2016,danielsen2021}.

In weakly crosslinked hydrogel surfaces, hydrodynamics plays a significant role in the tribological response of the system~\cite{simic2020}. For highly crosslinked surfaces, which are the focus of this work, the friction force is velocity independent or only weakly velocity dependent up to a threshold velocity~\cite{dunn2014,pitenis2014}. Past this threshold, experiments have shown that the friction coefficient increases approximately with the square root of the velocity for highly crosslinked surfaces~\cite{pitenis2014,uruena2015,meier2019,simic2020}.

Competing models exist as explanations for this behavior: Pitenis et al.~\cite{uruena2015,pitenis2018} argue that for hydrogels with  mesh size $\xi$, solvent viscosity $\eta$ and temperature $T$, friction is determined by a competition of fluctuation-induced polymer relaxation with time scale $\tau_\text{r}=\xi^3\eta/k_\text{B}T$ and the time scale introduced by the shear rate, $\tau_\text{s}=\xi/v$. This gives rise to velocity-independent (Coulomb) friction at small velocities $v$, where $\tau_\text{r}\ll\tau_\text{s}$. Friction measurements at high velocities collapse onto a single curve when plotted against the Weissenberg number~\cite{weissenberg1948}, $\mathrm{Wi} = \tau_\text{r}/\tau_\text{s}$. This velocity-dependent friction regime can thus be attributed to the non-equilibrium behavior of strained polymer chains and possibly adhesive interactions between the surfaces~\cite{pitenis2014}. Simple hydrodynamic-fluid-film friction models predict friction coefficients that scale with the square root of the velocity~\cite{hamrock2004}, but those models assume that surfaces are rigid and do not deform during sliding, which might not be representative of hydrogel-like biological surfaces. Furthermore, they require converging surfaces to generate hydrodynamic lift, a condition not met by parallel surfaces~\cite{mate2019}. In contrast, soft elasto-hydrodynamic lubrication theory~\cite{hamrock2004} allows the hydrogels to deform during sliding, but it also predicts friction coefficients that are lower than those measured experimentally and does not predict the scaling of friction with velocity observed in experiments~\cite{uruena2018}. These hydrodynamic lubrication theories also do not capture the transition between the velocity-independent and velocity-dependent friction regimes that appear to depend on the mesh size $\xi$~\cite{pitenis2014,uruena2015}. 

We present evidence from molecular simulations in favor of the polymer-relaxation hypothesis in highly crosslinked hydrogels. By choosing a model with an implicit solvent, i.e. without long-range hydrodynamic momentum transport and no possibility of interfacial fluid-film formation, we disentangle the role played by the solvent from that of the polymer network. We find that hydrogel friction is a purely interfacial phenomena at low contact pressures. Furthermore, the shear response of interfacial polymers can be split into three speed-dependent regimes. For small velocities, the friction coefficient is speed independent due to the effect of thermal fluctuations overwhelming chain alignment. This is followed by a regime in which the friction coefficient increases linearly with velocity. We trace this observation back to the reorientation of interfacial polymer chains in the direction of shear. Once forces are strong enough, these interfacial chains start stretching, which leads to a friction coefficient that increases with the square root of velocity. We complement our simulations with experiments that confirm that the friction coefficient increases with increasing mesh size for highly crosslinked systems. 

\section*{Results \& Discussion}
\begin{figure*}
\centering
\includegraphics{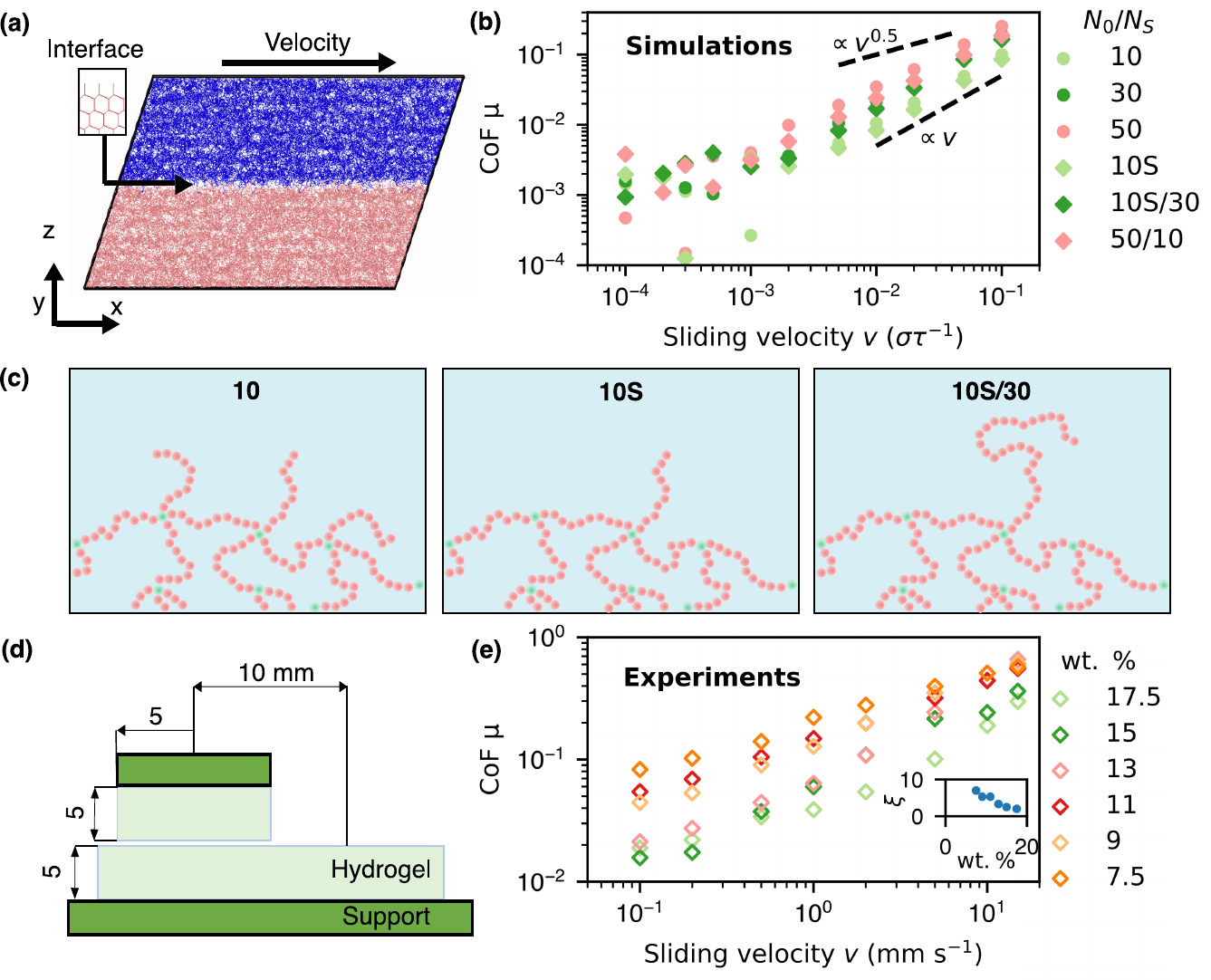}
\caption{ (a) Slice of the hydrogel network during shear. The network has been colored, in order to help distinguish the interface. The black line depicts the borders of the periodic simulation cell. The inset shows a detailed view of the interface before relaxation. (b) Coefficient of friction (CoF) as a function of sliding velocity $v$ and chain lengths for our computational model. (c) Sketch of the interface for networks with and without surface variations. Monomers are colored in red and crosslinkers in green. (d) Schematic of the friction experiments using a flat, hydrogel pin on a flat, highly crosslinked gel. (e) Experimentally determined coefficients of friction for polyacrylamide hydrogels as a function of monomer concentration. The inset shows estimated mesh sizes $\xi$ as a function of monomer weight percent.}
\label{fgr:friction}
\end{figure*}
Our molecular dynamics simulations~\cite{Allen1989-nt} resolve polymer-chain dynamics on length scales of the order of a mesh size (5-20~nm). We use a purely repulsive, flexible bead-spring model~\cite{kremer1990}, where a polymer chain is divided into uncorrelated segments (beads), each containing one or more monomers and of a size corresponding roughly to the Kuhn length~\cite{kuhn1934}. The polymer chains are thermalized by a dissipative particle dynamics thermostat, which acts as an implicit solvent, but mediates no long-range hydrodynamic interactions~\cite{espanol1995,hoogerbrugge1992}.

In order to construct our mesoscopic model, we mimic hydrogels used in friction experiments. These are disordered networks with four-fold coordination, physical crosslinkers, entanglements and potentially dangling chains. Unfortunately, the detailed statistical properties of the connectivity of hydrogel networks are often unknown~\cite{zhong2016,wang2018} and influence frictional behavior~\cite{meier2019,simic2020}. We therefore simplify our computational network and build the simplest regular structure with four-fold connectivity: a diamond lattice with dangling surface chains (see inset to Fig.~\ref{fgr:friction}a). Such diamond-like structures have been used in the past to test mechanical properties of polymer networks~\cite{everaers1995}. Four-fold coordinated crosslinkers are placed at the lattice sites and are connected to each other with polymer chains of predetermined length $N_0$.  In our model, a \{111\} crystallographic plane is parallel to the interface. As a result, crosslinkers at the surface are connected by three chains to the bulk and have one dangling chain of length $N_S$. Our canonical model uses identical lengths for bulk and surface dangling chains ($N_0=N_S$), but we vary these lengths independently to test their individual influence on the frictional properties. 

First, using our computational model, we look at the shear response of networks with the goal of unravelling the velocity dependence of the coefficient of friction. We do so by imposing shear on the polymer network by continuously deforming the periodic simulation cell at constant velocity $v$. The shear stress and normal stress are sampled every 10~time steps, for at least $10^8$ time steps, so as to compute the friction coefficient (CoF), $\mu=|\sigma_{xz}/\sigma_{zz}|$.  
We checked for select configurations that friction follows Amontons' law, i.e. the time-averaged shear stress $\sigma_{xz}$ is proportional to the normal stress $\sigma_{zz}$.

We observe two distinct regimes in the networks' shear response, as seen in Figure~\ref{fgr:friction}b. For all networks, above a certain threshold velocity, the friction coefficient increases approximately linearly with velocity. At low velocities, noise introduced by thermal fluctuations of the bulk chains prevent precise shear stress measurements. However, especially for networks with shorter chains, a change in friction behavior towards a speed-independent or only weakly speed dependent friction regime can be observed. Our simulations indicate that the threshold velocity between friction regimes is chain-length dependent: The threshold velocity is higher for networks with short chains than for networks with longer chains.

Next, we analyze the effect of surface chain density. Network $10$S is a network with $N=10$~beads/chain, in which dangling chains at the interface have been made sparse. (See our illustration of the near-surface network in Fig.~\ref{fgr:friction}c.) A lower surface chain-density slightly decreases the friction coefficient (Fig.~\ref{fgr:friction}b). To unravel the influence of bulk and surface chains, we vary the relative lengths of both. The network $10$S$/30$ has been constructed on the basis of network $10$S, but extending the dangling chains to $N_S=30$~beads/chain (Fig.~\ref{fgr:friction}c). In network $50/10$, the surface chains have been shortened to $N=10$~beads, while keeping the bulk network and chain density of $N_0=50$~beads. Figure~\ref{fgr:friction}b shows that both the transition velocity and magnitude of the friction coefficient are closest to the results of the network with $N_S~=~N_0=\max(N_0,N_S)$. Hence, the longest chains present at the interface dominate the frictional behavior.

We accompany our simulations by experiments on polyacrylamide gels with a highly crosslinked surface using a tribometer in reciprocating configuration (Fig.~\ref{fgr:friction}d), following the experiment design outlined in Meier et al.~\cite{meier2019}. Mesh sizes were extrapolated from Urue\~{n}a et al., but corrected with a factor of $0.15$~\cite{pitenis2018,pitenis_private}. Urue\~{n}a's data shows how for a constant monomer/crosslinker ratio, the mesh size decreases with increasing monomer concentration due to an increase in physical crosslinks (entanglements)~\cite{uruena2015}. We observe that the friction coefficient increases with sliding velocity (Fig.~\ref{fgr:friction}e). Furthermore, for any given sliding velocity, the friction coefficient increases with mesh size. This behavior is also observed in our simulations (Fig.~\ref{fgr:friction}b).  This last observation contradicts previous results by Urue\~{n}a et al., who show an inverse relationship between mesh size and coefficient of friction~\cite{uruena2015}. This may be a consequence of Urue\~{n}a et al. using a gel molded against polystyrene, whereas our gels were molded in glass petri dishes. The latter results in a more highly crosslinked surface~\cite{meier2019,simic2021}. Furthermore, the flat-on-flat tribometer configuration used in our experiments differs from the spherical indenter used by Urue\~{n}a et al., which may introduce additional hydrodynamic lift.

\begin{figure}
\centering
  \includegraphics{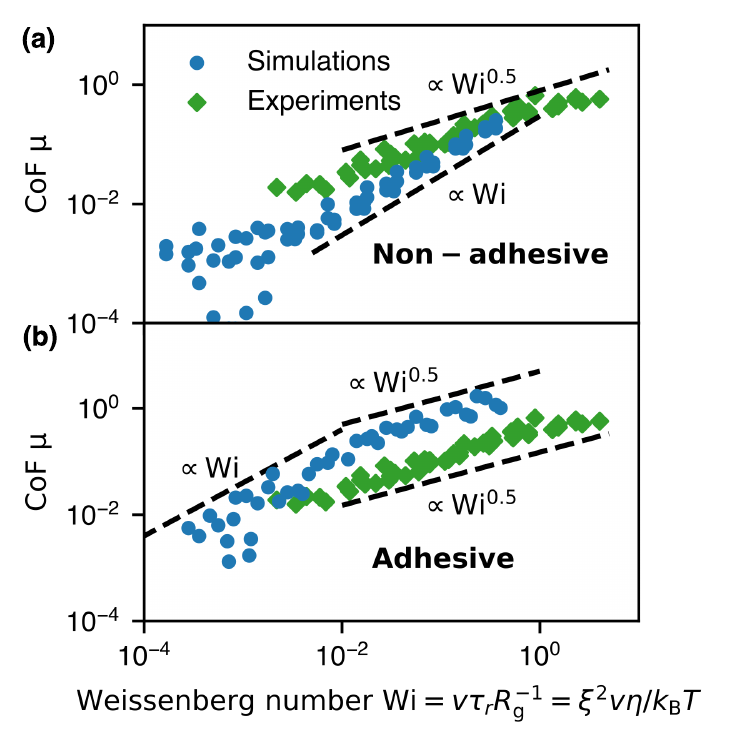}
  \caption{Coefficient of friction (CoF) as a  function of the Weissenberg number $\textrm{Wi}=v \tau_r R_{\rm{g}}^{-1}$ for the (a) non-adhesive and (b) adhesive models. Both are shown alongside experimental data for comparison.}
  \label{fgr:collapse}
\end{figure}

Following the experimentally observed collapse of the friction curves~\cite{uruena2015,pitenis2018}, we define the Weissenberg number as the product of the sliding velocity $v$ and polymer relaxation time $\tau_r$ divided by the radius of gyration $R_{\rm{g}}$ of a single chain of length $N_S$, $\mathrm{Wi}=v \tau_r R_{\rm{g}}^{-1}$ (see Supplemental Section S-I for information on how the relaxation time is computed for the single chain). This takes advantage of the fact that in the semi-dilute regime, the mesh size is approximately equal to the radius of gyration of a single chain~\cite{degennes1979}. For our experimental data we adopt the previously used definition of $\mathrm{Wi}=\xi^2 v \eta/k_{\rm{B}}\rm T$~\cite{uruena2015,pitenis2018}, with a relaxation time of $\tau_r\approx \eta\xi^3/k_{\rm B}T$~\cite{doi1988}. Figure~\ref{fgr:collapse}a shows a collapse of our friction curves onto experimental and computational master curves. In our simulations, the transition between friction regimes takes place close to $\mathrm{Wi}\approx 0.01$.
While both simulation and experiments show an increase in friction with velocity beyond this point, they differ qualitatively in the exponent.
The simulations show $\mu \propto \text{Wi}$ while the experiments scale as $\mu \propto \text{Wi}^{0.5}$.

A crucial simplification of our mesoscale simulation model is that the interfacial chains only interact repulsively through excluded volume interactions.
This type of model severely underestimates microscopic interchain friction and attractive interactions through hydrodynamic forces~\cite{chopra2002}, both phenomena that become relevant in nonequilibrium situations.
We here mimic such interactions by introducing a simple adhesive pair-interaction between both surfaces (see Methods for further details).
Figure~\ref{fgr:collapse}b shows that for the adhesive-interface model, the friction coefficient increases approximately as $\mu\propto\text{Wi}$ at small sliding speeds, but then crosses over to $\mu\propto\text{Wi}^{0.5}$, as in the experiment.
In this regime, the simulation and experimental coefficients of friction have the same exponent at high speeds. Experimental friction coefficients at high Weissenberg numbers are lower than friction coefficients from simulations. However, it is known that the exclusion of explicit solvent increases the macroscopic friction coefficient in polymer brush systems~\cite{galuschko2010}.  

The observed friction behavior is similar to what has been found in simulations for strongly compressed semi-dilute polymer brushes with explicit solvent~\cite{spirin2010,galuschko2010}. Given that sheared polymer brushes are subject to larger forces in explicit solvent compared to implicit solvent~\cite{galuschko2010}, our adhesive forces across the interface are mimicking the influence of an explicit solvent by increasing the shear forces acting on interfacial chains. This conjecture is supported by our results in the next section. 

In order to understand the molecular origins of the interfacial shear stress and the differences in behavior between the adhesive and non-adhesive models, we analyze the conformational statistics of chain segments at the hydrogel interface. In equilibrium, flexible chains take on random coil conformations. Polymer physics predicts that distorting equilibrium chain conformations reduces chain entropy and produces an entropic shear stress, defined as~\cite{oconnor2018,doi1988}
\begin{equation}
    \sigma_{xz}^{\rm entr} = \frac{\rho k_{\rm B}T}{C_\infty}\langle \mathcal{L}^{-1}(h)h \hat{r}_x \hat{r}_z \rangle.
    \label{eqn:entropy}
\end{equation}
Here, $\rho$ is the number density of monomers and $C_\infty$ is Flory's characteristic ratio, which gives the average number of monomer bonds in one random-walk step of the equilibrium coil conformation. For our model polymer chains $C_\infty = 1.83$~\cite{moreira2015}. The unit vector $\hat{\textbf{r}}$ points along the end-to-end distance vector $\vec{R}_{\rm{ee}}$ connecting the ends of a chain segment,  $\mathcal{L}^{-1}$ is the inverse Langevin function~\cite{rubinstein2003} and the extension ratio is defined as $h = |R_{\rm{ee}}|/nb$, with $n$ the number of bonds in a chain segment and $b$ the average monomer bond length.

\begin{figure}
\centering
  \includegraphics{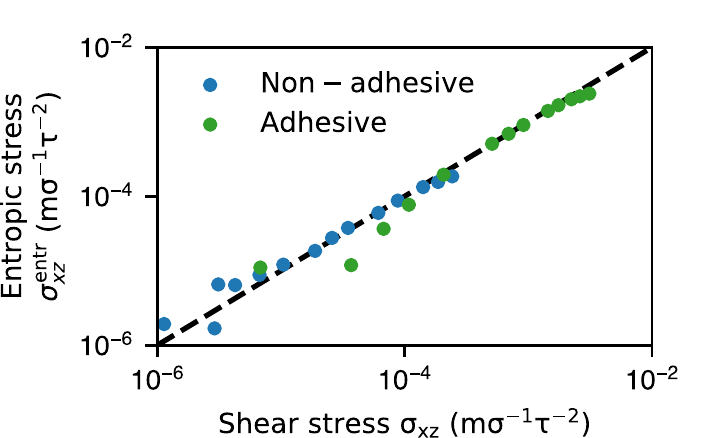}
  \caption{ Entropic stress as function of shear stress for the adhesive and non-adhesive models with $N=50$ beads/chain. Each data point corresponds to a different sliding velocity. }
  \label{fgr:entropicstress}
\end{figure}

To study the local shear stress of the interface, we restrict the ensemble average $\langle\cdot\rangle$ to dangling chain segments at the interface. Figure~\ref{fgr:entropicstress} plots this analytic entropic stress versus the total shear stress computed from the usual virial expression~\cite{Allen1989-nt} in our simulations. The two estimates agree, implying that the entropic stress of interfacial chains fully explains the shear stress for both the adhesive and non-adhesive models. Further, since the entropic prediction only included interfacial chains, this implies that bulk chains do not contribute substantially to the shear stress. They do not influence friction at low contact pressures. 

\begin{figure}
\centering
  \includegraphics{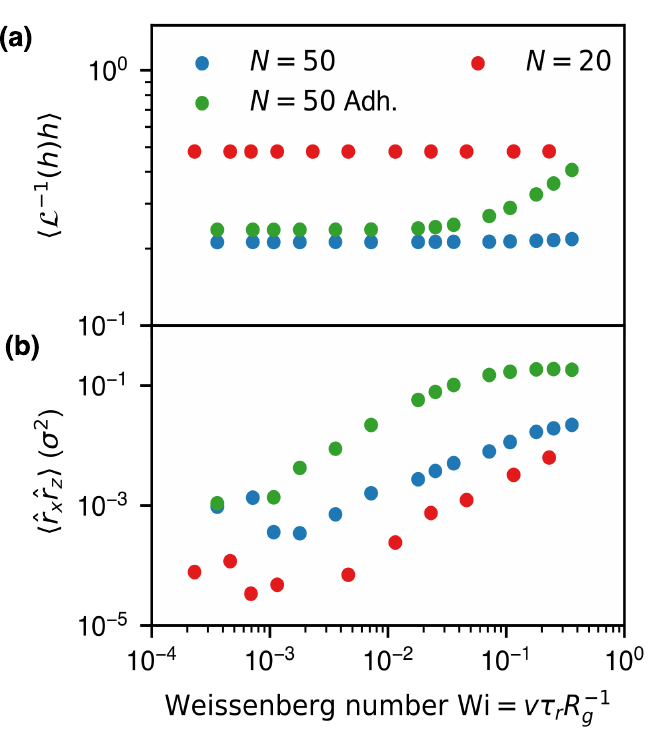}
  \caption{(a) Hardening factor of the entropic stress tensor as a function of the Weissenberg number for the adhesive ($N=50$) and non-adhesive models ($N=50$ and $N=20$). (b) Orientation tensor of the entropic stress tensor.}
  \label{fgr:entropic_wi}
\end{figure}

Since the entropic stress of chains captures the interfacial friction for the full range of $\rm{Wi}$, we can analyze it to understand the molecular processes governing the rate-dependent friction. Equation~\ref{eqn:entropy} can be decomposed into an orientation tensor $\hat{r}_\alpha \hat{r}_\beta$ that captures chain reorientation, and a Langevin hardening $\mathcal{L}^{-1}(h)h$ that captures the nonlinear effects of chain elongation. 
Figure~\ref{fgr:entropic_wi} plots these two terms separately for all systems  and $\rm{Wi}$. 
At low $\rm{Wi}$, chains relax faster than deformation distorts their conformations, and  show no change in orientation or elongation. This leads to a velocity-independent friction regime. As $\rm{Wi}$ increases, all systems show chain reorientation that grows with $\rm{Wi}$ as deformation tilts chain backbones towards the sliding plane. The adhesive system aligns more than the non-adhesive cases at the same $\rm{Wi}$, eventually saturating its alignment.
While all systems reorient, only the adhesive system shows any increase in chain elongation. Adhesive chains begin to elongate at larger $\rm{Wi}$, which coincides with the saturation in adhesive chain orientation. 

Hence, the shear stress is dominated by reorientation of interfacial chains for all systems at low sliding speeds (Fig.~\ref{fgr:entropic_wi}b).  Since a chain's relaxation time is proportional to its length, the velocity at which chains first noticeably reorient decreases with chain length. It follows that for any given velocity above the friction transition, longer chains should be further away from their equilibrium configurations than shorter chains (Fig.~\ref{fgr:entropic_wi}b).  As a consequence, the friction coefficient increases with chain length or mesh size (Fig~\ref{fgr:friction}b).

At low velocities, forces are small enough that interfacial chains relax faster than they are pulled out of equilibrium by the opposing interface. This is followed by the chains' reorientation along the shear direction, leading to a friction coefficient that increases linearly with velocity. Finally, interfacial chain stretching becomes significant, leading to a friction coefficient that is approximately proportional to the square root of velocity. Increased interfacial water content for gels with less crosslinked surfaces, or with a crosslinker gradient may change the dominant energy-dissipation mechanism~\cite{uruena2015,pitenis2018}.

We hypothesize that in the presence of explicit solvent, chain stretching would be more significant, even in the absence of additional adhesive interactions across the interface, since in the presence of explicit solvent, polymer chains are subject to larger shear forces than in equivalent implicit solvent simulations~\cite{galuschko2010}. The loss of molecular detail when adopting the mesoscopic bead-spring model might also lead to lower intermolecular friction. Therefore, we view the adhesive and purely repulsive networks as providing the shear response across the range of intermolecular force strengths.

\section*{Summary \& Conclusions}

In summary, we presented a coarse-grained computational model that reproduces the experimental phenomenology of self-mated hydrogel friction at low loads.
A speed-independent, Coulomb-friction regime at low sliding velocities crosses over to speed-dependent friction.
Simulation results quantitatively agree with experiments on hydrogels with a sharp gradient in surface density.
Our simulations reveal that the frictional response is solely dominated by interfacial chains, not by the bulk hydrogels.
The crossover to speed-dependent friction can then be attributed to the point where thermally driven relaxation of dangling polymer chains is slower than their excitation by sliding.
Our simulations further show that the velocity-dependent regime can be split into two stages: interfacial polymer chains first reorient along the shear direction and then stretch only once significant shear forces act on the chains.
These results show that tuning hydrogels for frictional properties should focus on engineering the surface, instead of the bulk structure.

\section*{Materials and methods}

\subsection*{Simulation methods}
Excluded volume interactions are modeled via a Weeks-Chandler-Anderson (WCA, cut-off radius $2^{1/6}\sigma$) potential~\cite{weeks_role_1971} with an interaction energy of $\varepsilon$ and a range $\sigma$, defining the time scale $\tau=\sigma (m/\varepsilon)^{1/2}$ of our simulations.  Consecutive segments in a polymer chain are connected by finite extensible nonlinear elastic (FENE) bonds~\cite{armstrong1974,kremer1990} with spring constant $30.0~\varepsilon \sigma^{-2}$ and a maximum extension $1.5~\sigma$ that prevents bond crossings. When adhesive interactions across the interface are included, we superimpose an adhesive Lennard-Jones potential of energy $\varepsilon_{\textrm{adh}}=0.5$, range $\sigma=1.0$ and cut-off radius $2\cdot2^{1/6}$ between dangling chains of each surface and bulk and dangling chains of the opposing surface. We report all simulation results in terms of the characteristic quantities $\sigma$, $\varepsilon$ and $\tau$. For friction simulations, we need a profile-unbiased thermostat and employ dissipative particle dynamics (DPD)~\cite{hoogerbrugge1992,espanol1995,soddemann_dissipative_2003,pastorino2007} with a linear weight function and a damping coefficient of $\gamma=5.0~m\tau^{-1}$. This thermostat is Galilean invariant and acts as an implicit solvent. Supplementary Section S-I contains a discussion of thermostat effects in this system. Equations of motion are integrated using a timestep of $\Delta t=0.001$.
Prior to running friction simulations, we equilibrate the networks using a Berendsen barostat~\cite{berendsen_molecular_1984}, such that our networks are stress-free parallel to the interface. We push both interfaces together by applying a pressure of $p_{zz}=-\sigma_{zz}=1\cdot10^{-3}~m\sigma^{-1}\tau^{-2}$ at a temperature of $T=1.0~\varepsilon$. The equilibration runs until the system size converges to a stable volume.  

\subsection*{Experimental methods}
The polyacrylamide (PAAm) hydrogels were synthesized from acrylamide monomer ($>$~99\%, Sigma-Aldrich, St. Louis, MO, USA), N,N’-methylene-bis-acrylamide cross-linker ($>$~99.5\%, Sigma-Aldrich, St. Louis, MO, USA) and 2,2'-azobis-(2-methylpropionamidine)dihydrochloride (98\%, Acros Organics, New Jersey) photo-initiator in MilliQ water. The molar ratio of the monomer to cross-linker was set to 50:1, where their total amount in the solution was set to 6, 7.5, 9, 11, 13, 15 and 17.5~wt.\%. The amount of the initiator was 0.5~wt.\%. Milli-Q water was bubbled with nitrogen for 30 minutes in order to remove the oxygen and then used to dissolve the set amounts of the monomer, the crosslinker and the initiator by gentle stirring. The solutions were poured into piranha-cleaned glass petri dishes to an approximate thickness of 3-4~mm and polymerized for 7~min by a UV light with an intensity of about 1~mW~cm$^{-2}$ at a wavelength of 365~nm (Stratalinker UV Crosslinker 2400, Stratagene Corp., La Jolla, CA, USA). After polymerization, the gels were removed from the molds and immersed in a large amount of Milli-Q water for at least 48~hours to remove unreacted species and allow swelling of the gels. Friction experiments were performed as described in Meier et al.~\cite{meier2019}.

\begin{acknowledgments}
The authors wish to thank Shiyun Tremp for synthesizing the hydrogels and performing the sliding experiments. The authors also thank Maryam Bahrami, Andrea Codrignani, Joachim Dzubiella, Johannes H\"ormann, Mark Robbins and J{\"u}rgen R{\"u}he for stimulating discussions. We used \texttt{LAMMPS}~\cite{plimpton_fast_1995} for all simulations and \texttt{Ovito~Pro}~\cite{stukowski2009} for visualization and post-processing. Computations were carried out on NEMO (University of Freiburg, DFG grant INST 39/963-1 FUGG), JURECA (Jülich Supercomputing Center, project \textit{hfr13}) and HOREKA (project \textit{HydroFriction}). We thank the Deutsche Forschungsgemeinschaft for providing funding within grants PA~2023/2 and EXC-2193/1 –- 390951807.

\end{acknowledgments}


%

\clearpage

\setcounter{figure}{0}
\setcounter{equation}{0}

\renewcommand{\thesection}{S-\Roman{section}}
\renewcommand{\thefigure}{S-\arabic{figure}}
\renewcommand{\thetable}{S-\arabic{table}}
\renewcommand{\theequation}{S-\arabic{equation}}

\begin{center}
\Large\bf{ Supplemental Material for\\"Molecular mechanisms of\\self-mated hydrogel friction" }
\end{center}

\section*{S-I Thermostating implicit solvent simulations}
\begin{figure}
\centering
  \includegraphics{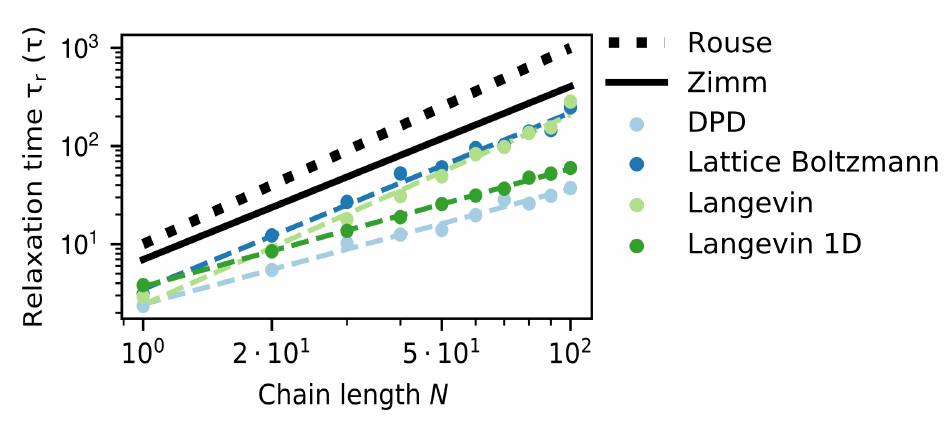}
  \caption{Polymer relaxation time $\tau_r$ as a function of the chain length $N$ for single chains and thermostat. Dashed lines represent fits. The black dotted line represents Rouse dynamics $\tau_r \propto N^{2}$ and the continuous black line represents Zimm dynamics with $\tau_r \propto N^{3\nu}$.}
  \label{fgr:relaxation}
\end{figure}

A polymer's relaxation time ($\tau_r$) can be computed from the autocorrelation function of the radius of gyration vector ($\textbf{R}_{\rm{g}}$)~\cite{doi1988}
\begin{equation}\label{eq:corr}
    \frac{\langle \textbf{R}_{\rm{g}}(t) \textbf{R}_{\rm{g}}(t_0) \rangle }{\langle \textbf{R}_{\rm{g}}^2(t_0) \rangle  } \propto \exp(-t/\tau_r)
\end{equation}
Figure~\ref{fgr:relaxation} shows the relation between the chain lengths and the polymer relaxation time for single chain systems. Results are shown for the DPD thermostat, the Langevin thermostat ($\gamma=0.5~m\tau^{-1}$) and explicit solvent  simulations using the fluctuating lattice Boltzmann method (FLB)~\cite{adhikari2005,dunweg2007}. We set the viscosity of the lattice Boltzmann fluid to $\eta=3.0~m\sigma^{-1}\tau^{-1}$ and the density equal to $\rho = 1.0~m\sigma^{-3}$. The FLB simulations used a timestep of $\Delta t= 0.01~\tau$ and lattice length of $\Delta x = 1.0~\sigma$. Each bead is coupled to the fluid by a force proportional to their relative velocities~\cite{ahlrichs1998}.
\begin{equation}
    \vec{F}=\gamma(\vec{u}-\vec{v_f})
\end{equation}
 $\vec{u}$ and $\vec{v_f}$ being the bead's and fluid's velocities, respectively. The coupling constant was set to $\gamma=20~m\tau^{-1}$. 
 
Regardless of the chosen thermostat, the radius of gyration scales with $R_{\rm{g}} \propto N^{\nu}$, the fitted Flory exponent $\nu=0.64$ being comparable to the theoretical value of $\nu=0.588$~\cite{degennes1979}.
The relaxation time scales with $N^{1.18}$ for the DPD thermostat, with $N^{1.80}$ for the FLB and with $N^{1.93}$ for the Langevin thermostat. As expected, the fluctuating lattice Boltzmann scaling is close to Zimm dynamics, while the Langevin thermostats scaling is close to Rouse dynamics. The DPD thermostat's scaling cannot be attributed to either theoretical model. In polymer systems, the DPD thermostat is usually restricted to medium-to-high densities, since in dilute regimes bead-solvent interactions play a significant role. Long range hydrodynamic interactions which give rise to Zimm dynamics rely on the presence of explicit fluid in the simulation, as is the case for our FLB simulations. However, FLB simulations, would have been computationally prohibitive to simulate large polymer networks. 

In practice, in order to circumvent its lack of Galilean invariance in non-equilibrium simulations, the Langevin thermostat is only used perpendicular to the direction of shear~\cite{thompson1990,pastorino2007}. Once we restrict the Langevin thermostat to a unidirectional thermalization, the relaxation time scales with $\tau_r \propto N^{1.20}$. Therefore, it shows similar dynamic scaling to the DPD thermostat. The DPD thermostat only includes central forces, so the question arises if Rouse dynamics can be achieved in dilute systems if the thermostat is extended to include non-central forces. The transverse DPD thermostat derived by Junghans et al. extends the standard DPD thermostat to thermalize the perpendicular components of the relative velocity between two beads, while still fulfilling the fluctuation dissipation theorem~\cite{junghans2008}. However, our results showed the relaxation time being proportional to $\tau_r \propto N^{1.14}$. Hence, we suspect that the DPD thermostat's failure to reproduce Rouse dynamics in the dilute regime is inherent to how it modifies the equations of motion. It is known that in the long wavelength limit, the DPD thermostat underdamps the dynamics, whereas the Langevin thermostat overdamps the dynamics~\cite{muser2006}.\\

\begin{figure}
\centering
  \includegraphics{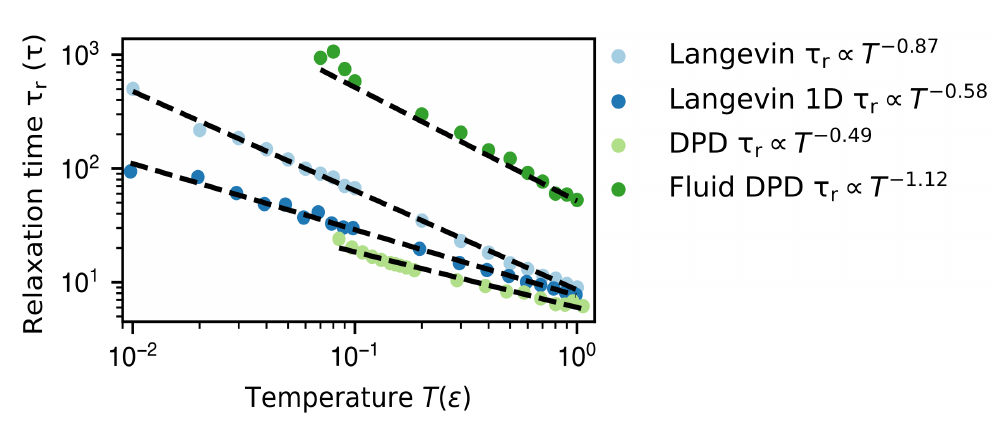}
  \caption{Polymer relaxation time $\tau_r$ as a function of temperature for a chain with 20~beads. Dashed lines represent fits.}
  \label{fgr:relaxation_temperature}
\end{figure}

Next, we study the temperature dependence of the relaxation time. Figure~\ref{fgr:relaxation_temperature} shows that a single chain relaxation with a Langevin thermostat is close to the theoretical result of $\tau_r \propto T^{-1}$. However, the scaling deviates from the theoretical prediction for single axis Langevin thermalization. Looking at the data for the DPD thermostat, we observe that in the single chain limit, the density is too low to successfully thermalize the lowest temperatures.  Furthermore, the relaxation time scales with approximately the square root of temperature. Once the chain is embedded in a repulsive Lennard-Jones (WCA) fluid of density 0.8, the relaxation time scales close to theoretical predictions for the DPD thermostat. We note that for all thermostats chosen here, the relaxation time decreases monotonically with increasing temperature.

We checked the influence of the thermostat on our stress measurements by comparing our DPD results to equivalent simulations using the one-dimensional Langevin thermostat. At high velocities there was a negligible difference between the shear stress observed between both. 
Simulations with explicit fluid were ruled out for our friction experiments with polymer networks due to computational cost. The one-dimensional Langevin thermostat failed to reproduce Rouse dynamics or the expected temperature scaling, as did the DPD thermostat. Therefore, we chose to use the standard DPD thermostat due to its built-in Galilean invariance and conservation of momentum. 

\end{document}